\documentclass{conm-p-l}

\copyrightinfo{2009}{}

\usepackage{graphicx}

\usepackage{amssymb,amsmath}

\usepackage{pstricks}
\usepackage{pst-plot}

\newtheorem{theorem}{Theorem}[section]

\theoremstyle{definition}

\theoremstyle{remark}

\numberwithin{equation}{section}

\newcommand{\e}{\epsilon}
\newcommand{\vth}{\vartheta}

\newcommand{\D}{{\mathcal D}}

\renewcommand{\k}{\kappa}
\newcommand{\ga}{\gamma}

\newcommand{\dl}{\delta}

\renewcommand{\th}{\theta}
\newcommand{\ra}{\rightarrow}
\newcommand{\al}{\alpha}
\newcommand{\be}{\beta}

\newcommand{\pa}{\partial}

\newcommand{\bq}{\bar{q}}

\newcommand{\hX}{\hat{X}}

\newcommand{\tU}{\tilde{U}}

\newcommand{\om}{\omega}

\newcommand{\htau}{\hat{\tau}}

\newcommand{\hvth}{\hat{\vartheta}}

\newcommand{\hH}{\hat{H}}

\begin{document}

\title[Issues of Chaos and Recurrence]
{Issues of Chaos and Recurrence in Infinite Dimensions}

\author{Y. Charles Li}

\address{Department of Mathematics, University of Missouri, 
Columbia, MO 65211}

\curraddr{}
\email{liyan@missouri.edu}

\thanks{}

\subjclass{Primary 35, 37; Secondary 76}
\date{}

\dedicatory{}


\begin{abstract}
Various issues with regard to chaos and recurrence in infinite dimensions are discussed.
The doctrine we are trying to derive is that Sobolev spaces over bounded spatial domains 
do host chaos and recurrence, while Sobolev spaces over unbounded spatial domains 
are lack of chaos and recurrence. Local Sobolev spaces over unbounded spatial domains 
can host chaos and are natural phase spaces e.g. for fluid problems, but are very 
challenging to study.
\end{abstract}

\maketitle








{\bf Many systems in applications are infinite dimensional systems. Infinite dimensional systems
are still not well studied or understood. Infinite dimensional dynamical systems have many 
novel features in contrast to finite dimensional dynamical systems, for example, different 
norms, and boundary conditions. Boundary conditions turn out to be very crucial to the dynamics, 
in particular whether or not the dynamics can be chaotic. For instance, decaying boundary conditions
over an unbounded spatial domain severely limit the development of chaotic dynamics. The commonly 
observed chaotic dynamics in nature, e.g. in fluid flows, does not obey any decaying boundary condition
at infinity. There are many ways to describe chaos. Here we utilize the description by its two 
phenotypes: sensitive dependence on initial data and recurrence. Using recurrence, we investigate 
what types of boundary conditions may support the existence of chaos. We find out that decaying 
boundary conditions over an unbounded spatial domain cannot support recurrence and chaos, while 
non-decaying boundary conditions or bounded spatial domains do support recurrence and chaos.}

\section{Introduction}

Chaos has two distinctive phenotypes: sensitive dependence on initial data, and 
recurrence. Sensitive dependence on initial data means that no matter how small 
the initial condition changes, after sufficiently long time, the change will reach order 
one. This phenotype can also be observed in non-chaotic systems, for example, in 
an explosive system. Together with the second phenotype, it can often identify 
chaos. Recurrence means that the orbit repeatedly re-visits the neighborhood of 
its initial point. 

Utilizing the phenotypes, one can often judge which dynamical system can host chaos 
and which cannot. In infinite dimensions, the dynamical systems are often defined 
by the Cauchy problems of partial differential equations. Here two key components 
influence the dynamics dramatically: the boundary condition and the norm of the 
phase space. As shown later on, periodic boundary condition (or other boundary conditions 
posed on a finite spatial domain) can often foster chaos, while decaying boundary 
condition (resulting in a Sobolev space on an infinite spatial domain) can hardly 
support chaos. For infinite spatial domain problems, relaxing the norm, say from 
a Sobolev space to a local Sobolev space, can bring chaos back into the larger phase 
space. But now the phase space is much more difficult to deal with.

\section{Lack of Recurrence for Unbounded Domain Problem}

Here we take the nonlinear Schr\"odinger equation (NLS) as the example, of course, the illustration 
is true for many other nonlinear wave equations. 
\begin{equation}
iq_t = q_{xx} +2|q|^2q ,
\label{NLS}
\end{equation}
where $q$ is a complex-valued function of two real variables ($t,x$). The NLS can be written in the 
Hamiltonian form
\[
-iq_t = \frac{\dl H}{\dl \bq} ,
\]
where 
\[
H = \int_{\D} [ |q_x|^2 - |q|^4 ] dx 
\]
and $\D$ is either $\mathbb{T}^1$ or $\mathbb{R}^1$, in either case, the NLS is globally well-posed 
in the Sobolev space $H^1(\D )$. 

For the unbounded domain $\mathbb{R}^1$, the natural phase space is for example 
a Sobolev space $H^1(\mathbb{R}^1)$. Now we have a Hamiltonian flow defined by the Cauchy problem 
of NLS in the phase space $H^1(\mathbb{R}^1)$. The natural question is: Is there recurrence?
Let us look at a traveling wave solution (say a soliton),
\[
q= Q(x-ct)\quad c \text{ is a constant.}
\]
It is clear that the Sobolev norm $\| Q \|_{H^s(\mathbb{R}^1)}$ ($s \geq 0$ integer) of the traveling 
wave solution is independent of time $t$. Thus the traveling wave solution travels on the surface 
of a sphere $S_*$ in $H^s(\mathbb{R}^1)$. It is also clear that as the time $t$ approaches infinity,
the traveling wave solution $Q(x-ct)$ has no limit in the phase space $H^1(\mathbb{R}^1)$. 
That is, the traveling wave solution $Q(x-ct)$ travels on the surface of the sphere $S_*$
one way and never returns --- no recurrence. On the other hand, in finite dimensional Hamiltonian 
systems, orbits on a sphere are always recurrent. 

Of course, there is a sequence of reasons to explain the new phenomenon in infinite dimensions. 
First, the finite dimensional invariant volume measure has no limit in infinite dimensions, therefore,
the finite dimensional recurrence argument of Poincar\'e is not valid in infinite dimensions 
\cite{Li09b}. Second, the Sobolev sphere $S_*$ in $H^s(\mathbb{R}^1)$ is not compact in $H^s(\mathbb{R}^1)$,
so that the compactness argument for recurrence fails \cite{Li09}. Third, the Sobolev sphere $S_*$ 
in $H^s(\mathbb{R}^1)$ ($s > 0$ integer) is not compactly 
embedded in $H^0(\mathbb{R}^1)$ since $\mathbb{R}^1$ is an unbounded spatial domain where the Rellich 
embedding fails, so that the argument of \cite{Li09} for recurrence in $H^0(\mathbb{R}^1)$ fails. 

In fact, general solutions of the NLS in $H^1(\mathbb{R}^1)$ are asymptotic 
to multiple soliton solutions of the form
\[
q= Q(x-c_1t, \cdots, x-c_nt)\quad c_j \text{ is a constant for } j = 1, \cdots , n;
\]
which also travel one way and never returns. Due to the lack of recurrence, one should not expect 
chaos in $H^1(\mathbb{R}^1)$  when the NLS is under perturbations. This claim should be true for 
general nonlinear wave equations in $H^s(\mathbb{R}^n)$.

\section{Recurrence for Bounded Spatial Domain Problem}

Now we turn to the bounded spatial domain problem, e.g. the periodic domain $\mathbb{T}^1$. 
We will show that the NLS flow is recurrent in $H^1(\mathbb{T}^1)$.
\begin{theorem}
For any $q \in H^1(\mathbb{T}^1)$, any $\dl >0$, and any $T>0$; there is a $q_* \in H^1(\mathbb{T}^1)$
such that 
\[
F^{n_j T} (q) \in B^0_\dl (q_*) = \{ q_1 \in H^1(\mathbb{T}^1) \ | \  \| q_1 - q_* \|_{L^2(\mathbb{T}^1)}  < \dl \}
\]
where $\{ n_j \}$ is an infinite sequence of positive integers, and $F^t$ is the evolution operator of the NLS. 
\label{RT1}
\end{theorem}
Before proving the theorem, we like to remark that the theorem roughly says that any $H^1(\mathbb{T}^1)$
solution to the NLS returns repeatedly to an arbitrarily small $L^2(\mathbb{T}^1)$ neighborhood. 
\begin{proof}
In order for the proof in \cite{Li09} to go through, we need to show that the $H^1(\mathbb{T}^1)$ norm 
is controlled by the two invariants of NLS, the Hamiltonian $H$ and the $L^2(\mathbb{T}^1)$ norm. 
Then any solution in $H^1(\mathbb{T}^1)$ shall stay in a bounded region $B$ in $H^1(\mathbb{T}^1)$ 
for all time. By the Rellich embedding theorem, $B$ is compactly embedded in $L^2(\mathbb{T}^1)$. 
Thus one can expect recurrence of the solution in $L^2(\mathbb{T}^1)$ as shown in \cite{Li09}.
By the Gagliardo-Nirenberg interpolation inequality \cite{Nir59}  \cite{Nir66},
\begin{equation}
\| q\|_{L^4} \leq C_1 \| q_x \|_{L^2}^{1/4} \| q \|_{L^2}^{3/4} + C_2 \| q \|_{L^2} .
\label{GN1}
\end{equation}
Using Young's inequality, one gets
\begin{eqnarray*}
\| q\|_{L^4}^4 &\leq& C ( \| q_x \|_{L^2}\| q \|^3_{L^2}+\| q \|_{L^2}^4)  \\
&\leq& C ( \k \| q_x \|_{L^2}^2 +\frac{1}{\k} \| q \|_{L^2}^6+\| q \|_{L^2}^4) \\
&=& \frac{1}{2} \| q_x \|_{L^2}^2 +2C^2  \| q \|_{L^2}^6+ C \| q \|_{L^2}^4,
\end{eqnarray*}
by choosing $\k = \frac{1}{2C}$. Thus, 
\[
\| q \|_{H^1}^2 \leq 2H +\| q \|_{L^2}^2 +4C^2  \| q \|_{L^2}^6+ 2C \| q \|_{L^2}^4.
\]
This completes the proof.
\end{proof}
For the unbounded domain $\mathbb{R}^1$, the $H^1(\mathbb{R}^1)$ norm of $q$ is still 
controlled by the two invariants of NLS, the Hamiltonian $H$ and the $L^2(\mathbb{R}^1)$ norm. 
In this case, $C_2=0$ in (\ref{GN1}). Unfortunately, as mentioned before, 
unlike the periodic domain $\mathbb{T}^1$ case, $H^1(\mathbb{R}^1)$ is not compactly 
embedded in $L^2(\mathbb{R}^1)$ due to the failure of Rellich's embedding theorem;
and the compactness argument in \cite{Li09} cannot be carried through. 

Next, we want to discuss the NLS under a Hamiltonian perturbation
\[
-iq_t =\frac{\dl \hH}{\dl \bq} , \quad \hH = H + H_1
\]
where $H_1$ is the perturbation. The same argument as in Theorem \ref{RT1} and in \cite{Li09} implies
the following theorem.
\begin{theorem}
If (a). the $L^2(\mathbb{T}^1)$ norm of $q$ is still an invariant for the perturbed NLS, (b). the 
$H^1(\mathbb{T}^1)$ norm of $q$ is controlled by $\hH$ and $\| q\|_{L^2(\mathbb{T}^1)}$;
then the recurrence theorem \ref{RT1} holds for the perturbed NLS.
\end{theorem}
A simple condition that guarantees the invariance of $\| q\|_{L^2(\mathbb{T}^1)}$ is: 
$\bq \frac{\dl H_1}{\dl \bq}$ is real.

\section{Local Sobolev Phase Spaces}

Sobolev space $H^s(\mathbb{R}^n)$ seems not able to host chaos. The decaying boundary 
condition at infinity limits the development of chaos. The natural next candidate to study is the 
local Sobolev space $H^s_{loc}(\mathbb{R}^n)$. Since it contains the subspace $H^s(\mathbb{T}^n)$,
$H^s_{loc}(\mathbb{R}^n)$ can certainly host chaos. On the other hand, $H^s_{loc}(\mathbb{R}^n)$ is 
a large space that is difficult to study. Usually there is no invariant manifold structure in it. One can often 
expects an invariant manifold structure in a Banach space with a countable base in which case the 
flow is equivalent to a system of infinite ordinary differential equations. Here $H^s_{loc}(\mathbb{R}^n)$ 
does not have a countable base. In fact, often the unstable, center and stable subspaces in 
$H^s_{loc}(\mathbb{R}^n)$  are not separated from each other \cite{Li05}. Even though it is challenging 
to study, $H^s_{loc}(\mathbb{R}^n)$ is often the natural phase space in applications, for example, in 
fluid dynamics. Take the plane Couette flow as the specific example. 
The plane Couette flow is governed by the Navier-Stokes equations
\begin{equation}
\pa_t u_i + u_j u_{i,j} = - p_{,i} +\e u_{i,jj} , \quad u_{i,i} = 0 ; 
\label{NS-Couette}
\end{equation}
where ($u_1,u_2,u_3$) are the three components of the fluid velocity along 
($x,y,z$) directions, $p$ is 
the pressure, and $\e = 1/R$ is the inverse of the Reynolds number. 
The boundary condition is
\begin{equation}
u_1(x, a, z) = \al , \quad u_1(x, b, z) = \be , \quad
u_j(x, a, z) = u_j(x, b, z) = 0, (j=2,3);
\label{BC-Couette}
\end{equation}
where $a<b$, $\al < \be$, and $u_i \ (i=1,2,3)$ are bounded in $x$ and $z$. In this case, the 
natural phase space (manifold) is 
\[
H^s_{loc}(\mathbb{R}^1 \times [a,b] \times \mathbb{R}^1)
\]
under the constraint of the boundary condition (\ref{BC-Couette}). Due to the difficulty in studying 
this phase space, current studies (both numerical and analytical) focus upon the restriction that 
$u$ is periodic in $x$ and $z$ in which case the spatial domain is bounded and the phase space (manifold) is simplified to
\[
H^s(\mathbb{T}^1 \times [a,b] \times \mathbb{T}^1)
\]
under the constraint of the boundary condition (\ref{BC-Couette}).

An interesting subspace of $H^s_{loc}(\mathbb{R}^1)$ is the space of spatially quasi-periodic 
functions:
\[
H^s_{quasi}(\mathbb{R}^1) = \{ q \in H^s_{loc}(\mathbb{R}^1) \ | \ q(x) = Q(\om_1x, \om_2x, 
\cdots , \om_nx) \} 
\]
where ($\om_1, \om_2, \cdots , \om_n$) is a quasi-periodic base. Even though it still does not 
have a countable base, the special nature of $H^s_{quasi}(\mathbb{R}^1)$ may make it easier 
to study. Again take the NLS (\ref{NLS}) as the example, one can obtain an explicit expression for 
a homoclinic orbit which is spatially quasi-periodic \cite{Li05}:
\begin{equation}
q(t,x)=\tilde{Q} + q_0(t) \sin \hvth_0 \Pi_2/\Pi_1 \ ,
\label{expf}
\end{equation}
where
\begin{equation}
q_0(t)=ae^{i\th(t)}\ , \quad \th(t)=-[2a^2t+ \ga ] \ ,
\label{upo}
\end{equation}
and $a$ is the amplitude and $\ga$ is the phase;
\begin{eqnarray*}
\tilde{Q}&=& q_0(t)[1+\sin \vth_0 \ \mbox{sech}\tau \cos X]^{-1} 
[\cos 2\vth_0 \\
& & - i \sin 2 \vth_0 \tanh \tau  -\sin \vth_0 
\ \mbox{sech}\tau \cos X]\ , \\
\Pi_1&=& \bigg [ (\sin \hvth_0)^2 (1+\sin \vth_0 \ \mbox{sech}\tau \cos X)^2
+\frac{1}{8} (\sin 2 \vth_0)^2(\mbox{sech}\tau)^2(1-\cos 2X) \bigg ] \\
& & (1+\sin \hvth_0 \ \mbox{sech}\htau \cos \hX )- \frac{1}{2}
\sin 2 \vth_0 \sin 2 \hvth_0\ \mbox{sech}\tau \ \mbox{sech}\htau \\
& & (1+\sin \vth_0 \ \mbox{sech}\tau \cos X)\sin X \sin \hX +
(\sin \vth_0 )^2 \bigg [ 1+2\sin \vth_0  \\
& & \mbox{sech}\tau \cos X + [(\cos X)^2-(\cos \vth_0 )^2](\mbox{sech}\tau )^2
\bigg ] \\
& & (1+\sin \hvth_0 \ \mbox{sech}\htau \cos \hX ) - 2 \sin \hvth_0
\sin \vth_0 \bigg [ \cos \hvth_0 \cos \vth_0 \\
& & \tanh \htau \tanh \tau +(\sin \vth_0 +\ \mbox{sech}\tau \cos X)
(\sin \hvth_0 + \ \mbox{sech}\htau \cos \hX )\bigg ] \\
& & (1+\sin \vth_0 \ \mbox{sech}\tau \cos X)\ , \\
\Pi_2&=& \bigg [ -2 (\sin \hvth_0)^2 (1+\sin \vth_0 \ \mbox{sech}\tau \cos X)^2
+\frac{1}{4} (\sin 2 \vth_0)^2(\mbox{sech}\tau)^2 \\
& & (1-\cos 2X) \bigg ](\sin \hvth_0 + \ \mbox{sech}\htau \cos \hX  + i 
\cos \hvth_0 \tanh \htau ) \\
& & + 2(\sin \vth_0 )^2(-\cos \vth_0 \tanh \tau + i \sin \vth_0  + i 
\ \mbox{sech}\tau \cos X)^2 \\
& & (\sin \hvth_0 + \ \mbox{sech}\htau \cos \hX -i \cos \hvth_0 \tanh \htau )
+2 \sin \vth_0  ( \sin \vth_0 \\
& & + \ \mbox{sech}\tau \cos X + i\cos \vth_0 \tanh \tau )[ 2 \sin \hvth_0 
(1+\sin \vth_0 \ \mbox{sech}\tau \cos X) \\
& & (1+\sin \hvth_0 \ \mbox{sech}\htau \cos \hX ) - \sin 2\vth_0
\cos \hvth_0\ \mbox{sech}\tau \ \mbox{sech}\htau \sin X \sin \hX ]\ , 
\end{eqnarray*}
and
\begin{eqnarray*}
& & \be_1 +i \sqrt{a^2-\be_1^2} = a e^{i\vth_0}\ , \quad \be_2 +i \sqrt{a^2-\be_2^2} = a e^{i\hvth_0}\ , \\
& & \tau = 4 \sqrt{a^2-\be_1^2}\be_1 t -\rho \ , \quad \htau = 4 \sqrt{a^2-\be_2^2} \be_2 t -
\hat{\rho} \ , \\
& & X= 2\be_1 x +\vth - \vth_0 +\pi /2 \ , \quad \hX= 2\be_2 x +\hvth - 
\hvth_0 +\pi /2 \ ;
\end{eqnarray*}
and $0< \be_1, \be_2 < a$, $a$, $\ga$, $\rho$, $\hat{\rho}$, $\vth$, and $\hvth$ are real 
parameters. 
As $t \ra \pm \infty$, 
\[
q(t,x) \ra q_0(t) e^{\mp i 2 (\vth_0 +\hvth_0 )} ,
\]
that is, $q(t,x)$ is homoclinic to the uniform periodic orbit $q_0(t)$ (\ref{upo}) up to 
phase translations. 
The unstable, center and stable subspaces in $H^s_{quasi}(\mathbb{R}^1)$  of the uniform 
periodic orbit $q_0(t)$ (\ref{upo}) are not separated from each other \cite{Li05}. The interesting 
question here is: Does the homoclinic 
orbit (\ref{expf}) induce chaos when the NLS is under perturbations ? 

\section{Temporal v.s. Spatial Evolutions}

From a dynamical system point of view, a natural evolution in infinite dimensions is a temporal one 
posed by the Cauchy problem of a partial differential equation. For such a Cauchy problem, looking 
at its spatial evolution is often awkward and ill-posed. Take the simple example: 
$u_t = u_{xx}$, $x \in \mathbb{R}$. 
For temporal evolution, one may choose the phase space to be the local space
$H^3_{loc}(\mathbb{R}^1)$. A 
simple orbit in $H^3_{loc}(\mathbb{R}^1)$ is
\[
u = u_0 e^{-t k^2}, \ u_0 =\cos kx, \ k \in \mathbb{R}^1.
\]
If this were a nonlinear equation, one could ask the question of sensitive dependence of $u$
on the initial data $u_0$, and the existence of temporal chaos. For all the orbits in 
$H^3_{loc}(\mathbb{R}^1)$, one can also look at their spatial evolution 
(development may be a better word). For example, one can rewrite the simple orbit as 
\[
u = u^0 \cos kx, \ u^0 =e^{-t k^2}, \ k \in \mathbb{R}^1.
\]
One can equip a temporal topology to these orbits, e.g. $C^0(\mathbb{R}^+)$ topology: 
$\| u \|_{C^0} = \sup_{t \in \mathbb{R}^+} | u(t)|$. Obviously
\[
u^0 =e^{-t k^2} \in C^0(\mathbb{R}^+), \text{ and } u \in C^0(\mathbb{R}^+) \text{ for any }
x \in \mathbb{R}.
\]
If this were a nonlinear equation, one could ask the question of sensitive dependence of $u$
on the initial data $u^0$ under the topology $C^0(\mathbb{R}^+)$, and the existence of spatial chaos.

\section{Tubular Chaos}

The two phenotypes of chaos can often be realized near a homoclinic orbit (or a heteroclinic cycle). When the stable and unstable 
manifolds of a saddle intersect transversally and form a tangle with a homoclinic orbit, recurrence can occur near the homoclinic 
orbit. The Lyapunov exponents near the homoclinic orbit will be positive due to the transversality of the intersection between 
the stable and unstable manifolds along the homoclinic orbit. The positive Lyapunov exponents lead to sensitive dependence 
upon initial data. 

In higher dimensions, it is natural to study homoclinic tubes (heteroclinically tubular cycles)  instead of homoclinic orbits 
(heteroclinic cycles). It turns out that like homoclinic orbits (heteroclinic cycles), homoclinic tubes  (heteroclinically 
tubular cycles) can also lead to chaos --- tubular chaos \cite{Li03a} \cite{Li03b} \cite{Li04c} \cite{Li06c}. Such a tubular chaos 
is represented by a Bernoulli shift dynamics on a Cantor set of submanifolds instead of points. These submanifolds can still 
contain finer scale chaos inside them, and continuing this process can lead to a chain of finer and finer scale chaos 
and form a chaos cascade \cite{Li04c}. The existence of tubular chaos shows that taking the averages of solutions with respect 
to a neighborhood of initial data will not eliminate the chaotic nature of the dynamics.

\section{Ubiquity of Fluid Instability}

One common feature of fluid flows is that it is easy for them to become unstable. Rarely there is a 
fluid flow that is stable for all values of its parameters. So there must be something universal 
(generic) that makes fluid flows unstable. The spatial oscillation $e^{ik \cdot x}$ could be the thing.
Since $e^{ik \cdot x}$ is a Fourier mode, it is certainly universal. Often the larger the $k$ is, 
the more unstable modes the oscillation $e^{ik \cdot x}$ induces. This will lead to more spatial 
disorder - more turbulent states. Next we mention some examples. 

For the 2D Kolmogorov flow (with a periodic boundary condition in each direction and an artificial 
force), spatially oscillatory shears ($\cos k y , 0$) are steady states. It turns out that the 
larger the $k$ is, the more unstable modes the shear has \cite{LLS04} \cite{Li05}. For the 2D plane 
Couette flow, even though the linear shear ($y,0$) is linearly stable for all values of the 
Reynolds number \cite{Rom73}, adding ``small'' spatial oscillations ($y+ \frac{A}{n} \sin (4n\pi y),
0$) will end up linearly unstable shears (they are steady for the inviscid flow, and slowly 
drifting for the viscous flow) for any $n$ and $A \in (\frac{1}{8\pi}, \frac{1}{4\pi})$ \cite{LL09}. 
The larger the $n$ is, the more unstable modes it has. 

Intuitively it is very easy to understand the destablizing effect of the small-amplitude high-frequency 
spatial oscillations. By the well-known Rayleigh criterion, a necessary condition for a shear $U(y)$ 
to have inviscid linear instability is that it has an inflection point. If a shear $U(y)$ does not have 
any inflection point, adding the small-amplitude high-frquency spatial oscillations can create inflection 
points even though such oscillations do not change the original velocity profile much. That is, the 
slight modification on the velocity profile $U(y)$,
\[
\tU (y) = U(y) + \frac{A}{n} \sin (ny) \quad \text{for large } n
\]
can have a significant modification on its second derivative $U''(y)$,
\[
\tU'' (y) = U''(y) - An \sin (ny) \quad \text{for large } n
\]
and creates a lot of inflection points and the potential of linear instabilities. Both shears $U(y)$ 
and $\tU (y)$ are steady states under the Euler dynamics. Under the Navier-Stokes dynamics, they may not 
be steady rather drift slowly in time. Such slowly drifting states can still play an important role 
in transient turbulence, 

In some sense, spatial oscillations are recurrent motions -- spatial shakings. Even small amplitude but
high frequency spatial oscillations can generate considerable vorticities. Derivatives of vorticities 
are even higher. Such structures not only make the fluid flows unstable, but also create a lot of 
unstable modes leading to more turbulent spatial disorders. Temporally oscillatory forcings are also 
known to generate instabilities of fluid flows, for example the well-known Faraday wave generation 
\cite{Far31} \cite{KG96} \cite{WBW03}.

\end{document}